\documentclass[aps,pre,twocolumn,showpacs,superscriptaddress,groupedaddress]{revtex4-1}
\usepackage{amsmath}
\usepackage{amssymb}
\usepackage{amsfonts}
\usepackage{graphicx}
\usepackage{dcolumn}
\usepackage{bm}
\usepackage{sidecap}
 \usepackage{subfloat}
\usepackage{parskip}
\usepackage{grffile}

\usepackage{color}
\usepackage{hyperref}
\usepackage{hhline}
\usepackage{mathtools}
\usepackage{graphics}
\usepackage{multirow}
\usepackage{verbatim}
\usepackage{longtable}
\usepackage{rotating}
\usepackage{setspace}
\usepackage{epsfig}
\usepackage{subfigure}
\usepackage{epstopdf}
\usepackage{gensymb}
\usepackage[normalem]{ulem}
\usepackage[table]{xcolor}

\makeatletter
\def\@eqnnum{{\normalsize \normalcolor (\theequation)}}
 \makeatother
 
\graphicspath{{./}{ER/main/}}

\begin{document}
\title{Optimized evolution of networks for principal eigenvector localization}
\author{Priodyuti Pradhan$^{1}$, Alok Yadav$^{1}$, Sanjiv K. Dwivedi$^{1}$ and Sarika Jalan$^{1,2}$\footnote{sarikajalan9@gmail.com}} 
\affiliation{1. Complex Systems Lab, Discipline of Physics, Indian Institute of Technology Indore, Khandwa Road, Simrol, Indore-453552, India} \affiliation{2. Centre for Biosciences and Biomedical Engineering, Indian Institute of Technology Indore, Khandwa Road, Simrol, Indore-453552, India}
\date{\today}

\begin{abstract}
Network science is increasingly being developed to get new insights about behavior and properties of complex systems represented in terms of nodes and interactions. One useful approach is investigating localization properties of eigenvectors having diverse applications including disease-spreading phenomena in underlying networks. In this work, we evolve an initial random network with an edge rewiring optimization technique considering the inverse participation ratio as a fitness function. The evolution process yields a network having localized principal eigenvector. We analyze various properties of the optimized networks and those obtained at the intermediate stage. Our investigations reveal the existence of few special structural features of such optimized networks including the presence of a set of edges which are necessary for the localization, and rewiring only one of them leads to a complete delocalization of the principal eigenvector. Our investigation reveals that PEV localization is not a consequence of a single network property, and preferably requires co-existence of various distinct structural as well as spectral features.
\end{abstract}

\pacs{89.75.Hc, 02.10.Yn, 5.40.-a}

 \maketitle
\section{Introduction}
Networks provide a simple framework to understand and predict properties of complex real-world systems by modeling them in terms of interacting units \cite{rev_net}.  This framework is particularly successful in explaining various mechanisms behind the emergence of collective behaviors of systems arising due to the local interaction patterns of their components. Principal eigenvector (PEV) corresponding to the maximum eigenvalue of the network's
adjacency matrix has been shown to be particularly helpful in getting insight into the propagation or localization of perturbation in the underlying systems \cite{Goltsev_prl2012}. One key factor of our interest is to understand properties of networks which may help in spreading or restricting perturbation in networks captured by PEV localization \cite{Allesina_nat2015}. For instance, during a disease outbreak, one will be interested in knowing if the disease will spread through the underlying network leading to the pandemic or will be localized to a smaller section of the network \cite{Goltsev_prl2012,metastable_2016}. Similarly, one may be interested in spreading a particular information, for instance, awareness of vaccination at the time of disease outbreak, or may wish to restrict or localize a perturbation like rumor propagation \cite{zanette_rumor_propagation2002}.

Furthermore, metal-insulator transition has been extensively studied using Anderson localization in solid-state physics \cite{LR1985} 
driving interest to investigate localization transition in complex networks \cite{localization_transition_net,localization_2009,clus_coef_ipr}. 
Localization of an eigenvector refers to a state when few components of the vector take very high values while rest of the components take small values. In the current study, we use inverse participation ratio (IPR) to quantify the eigenvector localization \cite{Goltsev_prl2012}. Localization properties of PEV have been shown to provide insight into the propagation of perturbation in mutualistic ecological networks \cite{Allesina_nat2015}, an existence of rare regions in brain networks \cite{brain_networks2013}, and efficient functioning of Google matrix \cite{dima_google2015}. Recently, eigenvector localizations properties have been related to scaling parameter of scale-free networks \cite{satorras_localization2016} as well as used 
for detecting communities in multilayer and temporal networks \cite{Taylor_comm_Multi2016}.

Goltsev {\it et al.} reported that PEV localization of scale-free networks exists only for the power law exponent being greater than a critical value \cite{Goltsev_prl2012}. On the contrary, Pastor-Satorras {\it et al.} have shown that PEV of all the power-law degree distributed networks are localized to some extent, with the existence of two different types of localization based on the degree of the nodes \cite{satorras_localization2016}. Nevertheless, they noted that these two different types of localization are not so evident in real-world networks \cite{satorras_localization2016}. Furthermore, localization has been investigated for eigenvector centrality defining the score of each node based on its neighborhood properties and is a common measure for determining the importance of nodes in networks. However, it was also found that the eigenvector centrality may fail upon consequence of PEV localization \cite{martin_evec_centrality_limitations2014}.
Network properties such as the presence of hubs, the existence of dense subgraph, and a 
power-law degree distribution are few factors known to make a network more localized than the corresponding random one \cite{martin_evec_centrality_limitations2014, Goltsev_prl2012}.
Another work mentioned that eigenvector localization positively correlates with the variance of weighted degree distribution \cite{Allesina_nat2015}. However, a merely presence or absence of these features may not guarantee a highly localized PEV. In other words, incorporating one of these features may not yield a network which has the most localized PEV for a given network size. 

All these insist a systematic investigation of the role of various network properties in the 
PEV localization as also suggested in \cite{satorras_localization2016}.  In the current study, we examine various structural and spectral properties of the networks when they are evolved from a delocalized to the localized state? We mainly concentrate on the following aspects:
How can one achieve a network having the most or highly localized PEV for a given network size? 
What are the particular structural properties or local patterns the network corresponding to the most localized PEV has? 

This article demonstrates that the most localized behavior of PEV does not have a direct 
correlation with a single structural property of the underlying network. The optimized network concerning localized PEV possesses a special structural feature highlighting requirement of several structural and spectral properties. Starting with a connected random network topology, we generate a network which has a highly localized PEV using an optimization technique for the network evolution. Rewiring of the network happens in such a manner that each rewiring step yields the PEV more localized than the previous step. We analyze various structural and spectral properties of the networks 
during the optimization process. Our analysis demonstrates that PEV localization is not a consequence of existence or absence an individual network property, rather requires combinations of many. We succeed in constructing a blueprint of the network topology corresponding to a highly localized PEV. Our investigations reveal that the most optimized network possesses a special structure. Furthermore, we find that there exists a set of edges in the optimized network, rewiring any one of them leads to a complete delocalization of the PEV. We show that this sensitivity of the PEV localization is related to the behavior of second largest eigenvalue of the underlying network. Our analysis further elucidates that there exists an evolution regime where networks are as localized as the optimized one, however, they are robust against single edge rewiring concerning PEV localization. 
Additionally, we learn that properties of the intermediate and the final optimized structure remain same irrespective of the topology of initial network structure.

\section{Methods}
We represent a simple undirected connected network as $\mathcal{G}=\{V,E\}$, where $V=\{v_1,v_2,\ldots v_N\}$ and $E=\{e_1,e_2,\ldots e_M\}$ represent the set of nodes and edges, where $N$ and $M$ denote the size of $V$ and $E$, respectively. The present work restricts to simple networks, i.e. the network without multiple connections and self-loop. We refer $E^c=\{e_1^c,e_2^c,\ldots,e_{(N(N-1)/2)-M}^c\}$ as the set of edges which are not present in $\mathcal{G}$. The adjacency matrix ($A$) corresponding to $\mathcal{G}$ is defined as, $a_{ij}=1$ if there is an edge between node $i$ and $j$, otherwise $a_{ij}=0$. Further, $d_{i}=\sum_{j=1}^N a_{ij}$ denotes the degree of node $v_i$ and $\{d_{i}\}_{i=1}^N$ stands for the degree sequence of $\mathcal{G}$. The spectrum of $\mathcal{G}$ is a set of eigenvalues $\{\lambda_1, \lambda_2, \ldots, \lambda_N\}$ of $A$ where  $\lambda_1 \geq \lambda_2 \geq \cdots \geq \lambda_N$, and corresponding eigenvectors are $\{X_1, X_2, \cdots, X_N\}$. As $A$ is a real symmetric matrix, all the eigenvalues are real. Moreover, $A$ is a nonnegative matrix and $\mathcal{G}$ is always connected, hence from the Perron-Frobenius theorem~\cite{newman2010}, all the entries of PEV are positive. Therefore, for a connected network, PEV is said to be localized when a large number 
of components take value near to zero and only a few have large values. We quantify the eigenvector localization using the IPR \cite{Goltsev_prl2012} as follows, 
\begin{equation} \label{eq_IPR}
IPR(X_{k})=\sum_{i=1}^N x_{i}^4 
\end{equation}
where $x_i$ is the $i^{th}$ component of normalized eigenvector, $X_k$ with $ k \in \{1,2,\ldots, N\}$, in the Euclidean norm.  
A delocalized eigenvector with component $[1/\sqrt{N},1/\sqrt{N},\ldots,1/\sqrt{N}]$ has an IPR value $1/N$, whereas a most localized eigenvector with components $[1,0,\ldots,0]$ leads to the IPR value equal to $1$. For a connected network, IPR value lies between these two extreme values. 

For a given $N$ and $M$, our aim is to get a connected network which has the most localized PEV corresponding to the maximum IPR value.
For a particular value of $N$ and $M$, if we can enumerate all the possible network configurations, the network corresponding to the 
maximum IPR value will be our desired network. The number of possible network configurations for a given $N$ and $M$ is of the order $\mathcal{O}(N^{2M})$ \cite{harary_graph_enumeration1973}.
Therefore, we formulate this problem through an optimization technique as follows.

Given an input graph $\mathcal{G}$ with $N$ vertices, $M$ edges and a function $\zeta:\Re^{N \times 1} \rightarrow \Re$, we want to compute the maximum possible value of $\zeta(\mathcal{G})$ over all the simple, connected, undirected, and unweighted graph $\mathcal{G}$. Thus, we are maximizing the objective function $\zeta(\mathcal{G})$=$IPR(X_1)=x_1^4 + \dots + x_N^4$ subject to the constraints that $x_1^2 + \dots + x_N^2 = 1$, and $0 < x_i < 1$. 
Furthermore, the optimization process helps us in assessing the impact of structural and 
spectral properties of networks on IPR value of PEV as networks
evolve from the delocalized to a localized state.
We refer the initial network as $\mathcal{G}_{init}$ and the optimized network as $\mathcal{G}_{opt}$. 

Starting from an initial connected random network, we use an edge rewiring approach based on a Monte Carlo algorithm to
obtain the most optimized network in an iterative manner. For a single edge rewiring process, we choose an edge $e_i \in E$ uniformly and independently at random from $\mathcal{G}$ and remove it. At the same time, we introduce an edge in the network from $E^c$, which preserves the total number of edges during the network evolution. The new network and the corresponding adjacency matrix are denoted as $\mathcal{G'}$ and $A'$, respectively. The eigenvalues and eigenvectors of $A'$ are indicated as $\{\lambda'_1, \lambda'_2, \ldots, \lambda'_N\}$ and $\{X'_1, X'_2, \cdots, X'_N\}$, respectively. It is important to remark that during the network evolution there is a possibility that an edge rewiring makes the network disconnected. We only approve those rewiring steps which yield a connected graph. We calculate the IPR value of PEV from $A$ and $A'$. If $IPR(X'_{1}) > 
IPR(X_{1})$, $A$ is replaced with $A'$. Therefore, in each time step, we get a network which has the PEV more localized than the previous network. We repeat the above steps until we obtain the maximum IPR value corresponding to the optimized network, $\mathcal{G}_{opt}$.
\begin{figure}[t]
\centering
\includegraphics[width=3.4in, height=1.8in]{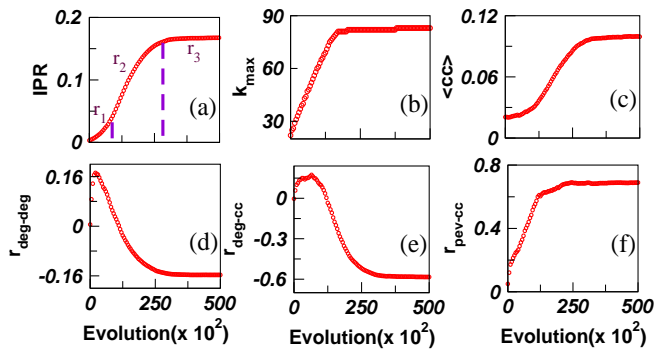}
\caption{Network size is $N=500$, $\langle k \rangle=10$ and we iterate the rewiring process for $600,000$ times and store the network after each $100^{th}$ steps. Changes of various network properties (a) IPR value of PEV, (b) maximum degree (c) average clustering coefficient, (d) degree-degree correlation, (e) correlation between degree vector and clustering coefficient vector, and (f) correlation between PEV and clustering coefficient vector during the evolution.}
\label{fig:fig_1}
\end{figure}
\section{Results}
Starting with an initial connected random network generated using Erd\"os-R\'enyi (ER) algorithm \cite{newman2010}, the evolution process based on the PEV localization forces a change in the initial network structure. The ER random network is generated with an edge probability $\langle k \rangle/N$, where $\langle k \rangle$ is the average degree of the network. 

\begin{figure}[t]
\centering
\includegraphics[width=3.4in, height=1.8in]{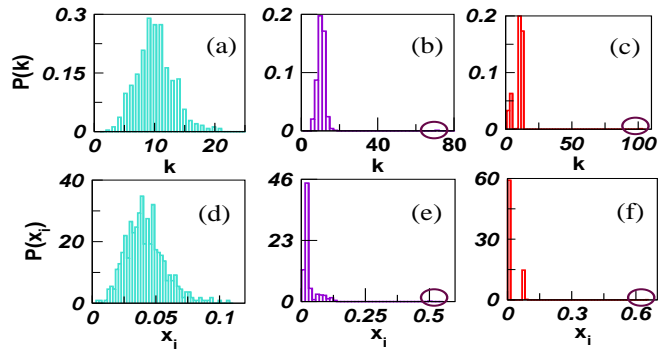}
\caption{ER network as initial network. Degree distribution of (a) initial (b) intermediate and (c) optimized network; PEV entry distribution of (d) initial (e) intermediate and (f) optimized network.}
\label{fig_degree_dist_ER}
\end{figure}

Based on the nature of changes in the IPR value, we can divide the evolution into three different regions; $r_1$, $r_2$, and $r_3$. In the first region, each rewiring yields a small change in the IPR value, whereas, in $r_2$ region, changes in the IPR values are much larger. The $r_3$ region
represents the saturation state (Fig.~1(a)). This is also referred as the critical region, explained later. At the beginning of the optimization process, the evolution of the IPR with rewiring is slow as there exist many nodes with degree close to each other (Fig.~\ref{fig_degree_dist_ER}(a)). Consequently,
for optimized rewiring, there exist several options for edges, rewiring which leads to an enhancement in the IPR value. Once, a node becomes clear hub by attaining considerable larger degree than the rest of the nodes (region $r_2$ and Fig.~\ref{fig_degree_dist_ER}(b)), the PEV entries corresponding to that node keeps on becoming larger (Fig.~\ref{fig_degree_dist_ER}(e)) and those of all other nodes get considerably smaller values, yielding a fast growth in the IPR values for each rewiring. 

As described in the previous section, the total number of edges is fixed throughout the evolution. Therefore, it is rearrangements of the links which affect the localization properties of the network. Moreover, we know that presence of localization affects many structural properties such as the largest degree ($k_{max}$),  the average clustering coefficient ($\langle CC \rangle$), degree distribution, etc.  \cite{martin_evec_centrality_limitations2014, Goltsev_prl2012, clus_coef_ipr}. We keep a record of all these properties during the evolution and observe that $k_{max}$ starts rising as IPR value increases (Fig.~\ref{fig:fig_1}(a)) and reaches its maximum value much before the IPR achieves its maxima (Fig.~\ref{fig:fig_1}(b)). This indicates that these two quantities affect each other positively, but they are not strongly related. Further to study a possible relation between $k_{max}$ and IPR value of PEV, we use the configuration model~\cite{genio_config_model2010} of the $\mathcal{G}_{opt}$ that has an IPR value which is much smaller than the optimal IPR value, even though both have the same $k_{max}$ and the degree sequence. This finding indicates that presence of a hub node or a particular degree sequence is important for PEV localization. Nevertheless, these may not be the only requirements for achieving a localized PEV, and thus we investigate other structural properties which contribute into the localization. 

The clustering coefficient is known to play a significant role in localization transition on complex networks~\cite{clus_coef_ipr}. 
We investigate $\langle CC \rangle$ vs. IPR during the evolution process. We find that as IPR value increases slowly in the $r_1$ region, 
while $\langle CC \rangle$ remains almost constant (Fig.~\ref{fig:fig_1}(c)). In the $r_2$ region, $\langle CC \rangle$ increases rapidly with the evolution and finally gets saturated to a particular value in $r_3$ region. It suggests that IPR and $\langle CC \rangle$ have a relation. One possible way to check this relationship is by constructing a network with the same $\{d_i\}_{i=1}^N$ and $\langle CC \rangle$ as for $\mathcal{G}_{opt}$ and to compare the IPR values of both the networks. Interestingly, the network constructed by the algorithm adopted from \cite{heath_cc_tuning2011}
has IPR value far lesser than the $\mathcal{G}_{opt}$. This experiment indicates that regulating $\langle CC \rangle$ leads to a localization of PEV, but it is not as high as $\mathcal{G}_{opt}$. Therefore, we investigate other structural properties which might contribute to the PEV localization. One such property is the degree-degree correlation of the networks which we measure using Pearson product-moment correlation coefficient \cite{newman2010}. 
\begin{figure}[t]
\centering
\includegraphics[width=0.8\columnwidth]{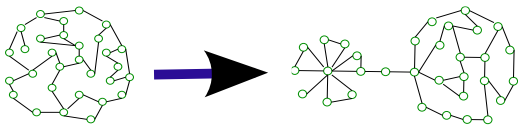}
\caption{(Color Online) Schematic diagram representing the initial (left)
and the most optimized (right)
networks.}
\label{fig_shematic}
\end{figure}

The degree-degree correlation ($r_{deg-deg}$) during the evolution process exhibits an increment in the beginning when there is a small change in the IPR value ($r_1$) and decreases rapidly with a further increase in IPR value ($r_2$). Finally, both become saturated (Fig.~\ref{fig:fig_1}(d)) 
and $\mathcal{G}_{opt}$ is a disassortative network. To check the importance of disassortativity for the localization, we perform an experiment by constructing a network using Sokolov algorithm~\cite{brunet_sokolov_algo2004} which has the same $r_{deg-deg}$ as of the $\mathcal{G}_{opt}$. 
However, this construction also fails to yield the IPR value as high as for $\mathcal{G}_{opt}$. Further, the degree and clustering coefficient vectors manifest a negative correlation (Fig.~\ref{fig:fig_1}(e)) whereas local clustering coefficient and PEV indicates a high positive correlation (Fig.~\ref{fig:fig_1}(f)). These two measurements do provide us information about the possible structure of the networks but do not tell what the structure exactly is. 

The most intriguing result of our investigation is that we get a special network topology corresponding to the optimized IPR value concerning PEV. The optimized network consists of two components of different sizes which are connected to each other via a single node (Fig.~\ref{fig_shematic}). 
In the beginning of the evolution process (starting at $r_1$ region), the degree distribution of the network follows Poisson law (Fig.~\ref{fig_degree_dist_ER} (a)).  
The evolution process forces to change the network structure in a very typical manner such that the degree of one node becomes much higher than the rest of the nodes in the network at the intermediate stage (Fig.~\ref{fig_degree_dist_ER}(b)). 
In the $r_3$ region, the degree distribution of the optimal structure which has the most localized PEV is depicted in Fig.~\ref{fig_degree_dist_ER}(c). One can notice that it has two peaks at lower $k$ values, and there exists one point corresponding to the hub node lying very far from these two peaks. The first smaller peak is contributed by the nodes lying in the smaller part of the network (Fig.~\ref{fig_shematic}(left)), and the larger peak is coming from, the larger component having optimized network structure. 
Similarly, the distribution of the PEV entries during the network evolution take shape
in such a manner that at the $r_2$ and $r_3$ regions (in Fig.~\ref{fig_degree_dist_ER}(e) and 
Fig.~\ref{fig_degree_dist_ER}(f)) more number of nodes have tiny weights at corresponding PEV
entries and less 
number of nodes have large weight which is an indication of a highly localized PEV 
(Fig.~\ref{fig_degree_dist_ER}(d)). Further, it is visible in Fig.~\ref{fig_degree_dist_ER}(f) that each node belonging to 
the small component (Fig.~\ref{fig_shematic}(left)) of the $\mathcal{G}_{opt}$ has large 
PEV weight whereas those belonging to the larger component has smaller PEV weights.
\begin{figure}[t]
\centering
\includegraphics[width=3.4in, height=1.6in]{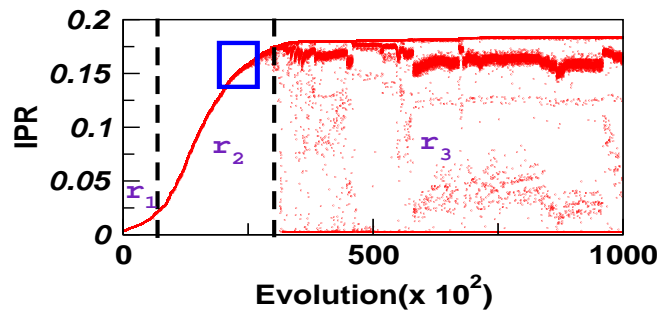}
\caption{IPR as a function of edge-rewiring. The networks with large IPR value in $r_3$ region consists of few edge-rewiring, which leads to a sudden drop in the IPR value. Rewiring of the first 1,00,000 edges is depicted. The marked square indicates the regime where the networks attain IPR values which are very close to the optimized network. However, in this regime rewiring an edge does not have a significant impact on the IPR values.}
\label{fig:fig_2}
\end{figure}

Note that, in between any two increments in IPR values as evolution progresses, there exists several edge rewiring which does not lead to an increase in the IPR value.  If we consider rewiring of all the edges, and not only those which lead to an increase in the IPR value, we get surprising results. In the $r_3$ region (Fig.~\ref{fig:fig_2}), IPR value gets almost saturated, and there exists only a subtle increment in its value with a further evolution of the network. Though the network in this region has the maximum IPR value, there exist few edges, rewiring them leads to a sudden drop in the IPR value resulting in the complete delocalization of PEV from  a highly localized state. It reveals that only a single edge rewiring makes the most localized PEV to delocalized and this phenomenon is observed for sparse networks in $r_3$ region (Fig.~\ref{fig:fig_2}). We look forward to identifying the set of special edges and the rewiring locations, perturbing which, lead to delocalization of PEV. It turns out that in the optimized network if we remove an edge connected to the hub node inside the smaller component (Fig.~\ref{fig_shematic}(right)) the IPR value drops down leading to a complete delocalization of PEV. Interestingly, just before the saturation (region $r_2$) if we rewire an edge which is connected to the hub, no sudden drop is observed in the IPR value. This is a region highlighted within a square in (Fig.~\ref{fig:fig_2}) where IPR value is much larger than the initial ER random network as well as is robust against the edge rewiring. Whereas in the $r_3$ region, though the network achieves the maximum IPR value, it becomes very sensitive to the single edge rewiring. Most importantly, by controlling few edges, we can control the PEV localization of the entire network. 

To check the robustness of the emerged localized network structure against changes in the initial network, we start the 
evolution process on the scale-free (SF) network considered as the initial structure. The SF network is constructed using Barabasi-Albert preferential attachment model \cite{newman2010}. The network gets evolved through the similar $r_1$, $r_2$ region of slow and fast changes in IPR values, and finally, leads to the saturation region $r_3$. The final optimal structure remains same as depicted by Fig.~\ref{fig_shematic}. There exist few changes occurring before the network reaches to the final optimized structure. A prime change is that reaching to the saturation state ($r_3$) is faster when one starts with an SF network structure. The reason behind this slightly faster convergence is that the PEV of the SF network is already slightly localized due to the presence of a hub node. Moreover, the optimization process acts on a network already having a hub node 
which causes shrinkage in the slow evolution region ($r_1$).
\begin{figure}[t]
\centering
\includegraphics[width=2.9in, height=1in]{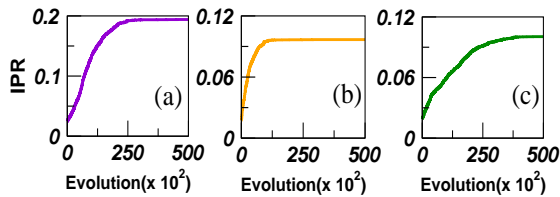}
\caption{Change in IPR value as a function of network evolution. (a) SF networks with $N=500$ and
$\langle k \rangle =10$ (b) C.elegans frontal network with $N=131$ and $\langle k \rangle =11$
(c) C.elegans neural network with $N=297$ and $\langle k \rangle =14$. We consider crude approximations that edges are undirected and unweighted for C.elegans networks.}
\label{SF_Celegans_IPR}
\end{figure}

\begin{figure}[t]
\centering
\includegraphics[width=2.4in, height=1.1in]{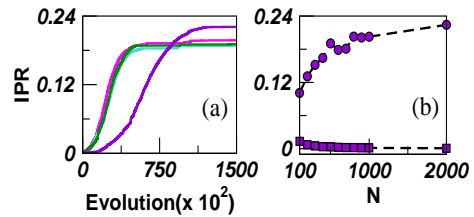}
\caption{(a) Evolution of IPR for three different network size. As network size increases, it takes more evolution time for a network to reach to the optimized state. (b) IPR value of the initial ($\blacksquare$) ER and the final ($\bullet$) optimized networks as a function of $N$.}
\label{N_ipr}
\end{figure}

Additionally, we consider few real-world networks as our starting initial network structure, and again the final optimal network structure 
remains the same as found earlier with the existence of critical region $r_3$. For example, we consider C.elegans frontal \cite{celegans_frontal_2006} and C.elegans neural \cite{celegans_neural_1998} network as the initial network structure and achieve the similar structure as obtained from the ER and SF networks through the evolution process (Fig.~\ref{SF_Celegans_IPR}). Further, we consider the impact of changes in the network size on the properties of the optimized network structure. As network size increases, the evolution process remains same as depicted by Fig.~\ref{N_ipr}(a). The final optimized network structure achieves through the intermediate stage and attains the same structure (Fig.~\ref{fig_shematic}). However, as $N$ increases, it takes more evolution time for a network to be optimized (Fig.~\ref{N_ipr}(a)). It is not surprising as  Goltsev et al. have provided theoretical bounds on the maximum IPR values for the Bethe lattices and have shown its dependency on $N$ \cite{Goltsev_prl2012}. Fig.~\ref{N_ipr}(b) depicts IPR values of the initial and the optimized network for various values of $N$. 

\begin{figure}[t]
\centering
\includegraphics[width=1\columnwidth]{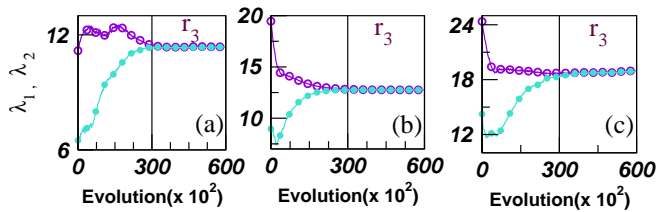}
\caption{Behavior of $\lambda_1$ ($\circ$)  and $\lambda_2$ ($\bullet$) during the network evolution. Initial network as (a) ER random network (b) SF network (c) C.elegans neural network.}
\label{lambda}
\end{figure}

In the following, we attempt to understand the emergence of the special structure as a consequence of optimization as well as the sensitivity of PEV in the critical region. The eigenvalues of a network adjacency matrix lie in a bulk region separated from extremal eigenvalues at both the side which lie outside the bulk. It is known that the extremal eigenvalues, particularly the largest one, may follow completely different statistical properties than those lie in the bulk \cite{SJ_Sanjiv_pre_gev_2014}. Furthermore, the number of eigenvalues lying outside the bulk is known to be equal to the number of communities in the network \cite{eig_community}. For a random network without any community structure, there exists only one eigenvalue which lies outside the bulk region and all other eigenvalues including the second largest $\lambda_2$ are part of the bulk region \cite{eig_community}. As depicted in Fig.~\ref{lambda}, value of $\lambda_2$ is much smaller than the value of 
$\lambda_1$ in the initial network structure corresponding to ER, SF, and C.elegans neural 
networks. During the evolution, $\lambda_2$ starts shifting towards $\lambda_1$, i.e. $\lambda_2$ starts drifting away from the bulk region. This drift in $\lambda_2$ is not surprising as we know that the final optimized structure consists of two parts or communities, and hence there should be two eigenvalues which lie outside the bulk. However, the interesting observation is that for the optimized network, $\lambda_2$ not only drift away from the bulk but becomes very close to $\lambda_1$, in fact, $\lambda_1 \sim \lambda_2$. 
Almost the same values for both the eigenvalues might be a reason behind sensitivity of PEV for a single edge rewiring. Markov chain and its associated transition matrix have been extensively studied in network science. It has been reported that when the two largest eigenvalues of a transition matrix become very close to each other, PEV which is known as the stationary probability distribution vector becomes sensitive to a small perturbation in the transition matrix \cite{sensitivity_markov_chain_1994}. Consequently, the associated Markov chain becomes decomposable \cite{subdominant_eigenval_1998}. The transition matrices are different from the adjacency matrices considered here; nevertheless, largest two eigenvalues being close to each other and sensitivity of PEV occurring at the same evolution time is brings forward an insight into the behavior of PEV localization.  When PEV becomes highly localized resulting in $\lambda_1 \sim \lambda_2$, the network structure becomes very sensitive for rewiring and may lead to a complete delocalization of PEV even for a single edge rewiring (Fig.~\ref{fig:fig_2}). Nevertheless, there exists open question that why the network having a highly localized PEV displays $\lambda_1 \sim \lambda_2$. This finding brings about more physics in the context of PEV localization and invites further investigations and analytical treatment.

\section{Disease spreading in localized networks} 
\begin{figure}[t]
\centering
\includegraphics[width=0.5\columnwidth]{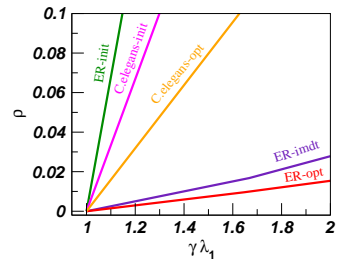}
\caption{Spreading process of SIS model on the initial ER random network ($\lambda^{init}_{1} \approx 11.34$, $IPR_{init} \approx 0.0007$), intermediate ($\lambda^{imdt}_{1} \approx 11.14$, $IPR_{imdt} \approx 0.21$) and optimized networks ($\lambda^{opt}_{1} \approx 10.77$, $IPR_{opt} \approx 0.22$) and the C.elegans neural networks ($\lambda^{init}_{1} \approx 24.36$, $IPR_{init} \approx 0.019$ and $\lambda^{opt}_{1} \approx 17.18$, $IPR_{opt} \approx 0.1$) have been depicted. ER network has $N = 2000$ nodes with $\langle k \rangle =10$.}
\label{fig:fig_4}
\end{figure}

Starting from the ER random or SF networks as an initial network, we can achieve an optimized network structure which has highly localized PEV. 
To demonstrate the efficiency of these artificially constructed network structures for a dynamic process, we use the standard susceptible-infected-susceptible (SIS) disease spreading model \cite{Goltsev_prl2012}. We observe the behavior of spreading process at different stages of the optimization process. In the SIS model, each susceptible vertex becomes infected with the infection rate $\gamma$ and infected vertices become susceptible with the unit rate. With probability $\rho_{i}$, a vertex $i$ infected by its neighbours, and the prevalence is given by $\rho = \sum_{i=1}^N \rho_{i}/N$. We know that when the infection rate $\gamma$ cross the epidemic threshold i.e. $\gamma > \gamma_c=\frac{1}{\lambda_1}$ \cite{epidemic_threshold2003} the disease will spread over the networks. However, if the PEV of the network's adjacency matrix is localized, in the vicinity of the epidemic threshold $\gamma_c + \epsilon, \; \epsilon>0$ the disease infects a small number of vertices and spreading process becomes slow. As a result, it requires 
a larger value of $\gamma$ for spreading the disease over the network. Fig.~\ref{fig:fig_4} manifests that for the initial network, for a value of $\gamma$ which is slightly larger than  $\gamma_c$, disease infects a large number of vertices. Whereas, for the networks corresponding to the intermediate and the optimized states, there exist very few vertices which get infected. 
\section{Conclusion}
In this work, starting from an initial random network, we achieve a network structure through a Monte Carlo based optimization method. The final optimized network possesses a highly localized PEV quantified by the IPR value. We analyze various structural and spectral properties of the optimized network as well as the networks at the intermediate state before the optimized structure is reached. In other words, we develop a learning framework to explore localization of eigenvector through a sampling-based optimization method. We demonstrate that PEV localization is not a consequence of a single network property and rather requires co-existence of several structural features. The final optimized network possesses a special structure and which we have shown to be robust against changes in the initial network structure. We demonstrate the robustness of the results by considering various popular network models as well as two real-world networks as an initial network structure. Furthermore, we characterize the evolution regime into different states. In the intermediate state, though the networks are not best optimized in terms of the PEV localization, they are robust against changes in single edge rewiring. Whereas, PEV is sensitive against single edge rewiring
in the critical region. Our analysis identifies a special set of edges which are essential for 
the (de)localization of PEV in the most optimized network structure. Rewiring any one edge of 
this set leads to a complete delocalization of PEV. We observe that this emergence of sensitivity in the PEV and shifting of $\lambda_2$ close to $\lambda_1$ happens simultaneously suggesting a relation between the special structure of the optimized network and the second largest eigenvalue, in addition to the first, of the network. We further identify the evolution regime which corresponds to the networks having PEV localization almost same as that of the optimized network but the localization property being robust against the edge rewiring. 
It may not always be feasible to rewire a real-world network to such an extent so as to get a desired PEV localization behavior. However, the results and approach here will be more useful in constructing an artificial network with a desired localization behavior. Further, there exist few real-world systems where rewiring can be achieved rather easily. For instance, functional connectivity in the brain can be changed by providing suitable input to get a desired localization behavior \cite{Diessen_Epilepsia_2013}. This is particularly relevant for epileptic seizure as PEV localization has been shown to be useful to understand criticality in brain networks \cite{brain_networks2013}.

Eigenvector localization is an important aspect of potential use in understanding propagation mechanisms in various systems such as the virus spread in computer networks \cite{computer_virus2004}, vibration confinement systems like spring-mass-damping systems and for constructing piezoelectric networks \cite{SC1996TW2004,passive}. Further, the present work focuses on undirected and unweighted networks; the approach can be extended to obtain a comprehensive picture of PEV localization on directed and weighted networks. Furthermore, this paper is restricted to PEV of adjacency matrices; however, it will be interesting to study the consequence of other lower order eigenvectors localization on emerging network properties \cite{cucuringu_low_order_evec2013}.

To conclude, our study provides a deeper insight to the PEV localization on synthetic as well as on empirical networks. Though, the prime concern of our analysis to have insights into the network structure and PEV localization, using SIS model, we verify that in the optimal and the intermediate stages spreading of disease is much slower than the initial random structure. Additionally, earlier work has related spectral properties with change in the matrix elements \cite{Meighem_pre2011}; here we show that how a function of PEV relates with the change in the matrix elements arising due to the edge rewiring. Moreover, the identification of critical region where a 
sudden IPR drops happen adds another dimension to eigenvector behavior of complex networks.

\begin{acknowledgments} 
SJ and AY, respectively acknowledge DST, Govt of India grant EMR/2014/000368 and CSIR 09/1022(0013)/2014-EMR-I
for financial support.  We are thankful to J. N. Bandyopadhyay (BITS-Pilani) for stimulating discussions on eigenvector localizations. We are 
indebted to Charo I. del Genio (University of Warwick) for several fruitful comments, particularly those related to optimization algorithm
 and Manavendra Mahato (IIT Indore), Complex Systems Lab members and Jan Nagler (ETH Zurich) for several useful suggestions.
\end{acknowledgments}

\end{document}